\documentclass[10pt]{article}
\usepackage{epsfig}
\usepackage{amsfonts}
\newcommand{\be}{\begin{equation}}   
\newcommand{\ee}{\end{equation}}   
\newcommand{\beqn}{\begin{eqnarray}}   
\newcommand{\eeqn}{\end{eqnarray}}
\begin{document}

\begin{center}       
\begin{picture}(0.2,0.2)
\end{picture} \\
{\Large\bf Massive c${\bf\bar c}$g-Production in Diffractive DIS}
\\[1cm]
J.Bartels$^1$,  H.Jung$^2$, and A.Kyrieleis$^1$
\\[0.7cm]
{\small\sl 
$^1$II. Institut f\"ur Theoretische Physik; Universit\"at Hamburg\\ 
Luruper Chaussee 149, 22761 Hamburg, Germany \footnote{Supported by the 
TMR Network ``QCD and Deep Structure of Elementary Particles''}\\
$^2$Physics Department, Lund University, P.O Box 118, 
221 00 Lund, Sweden}
\\[1.8cm]
\begin{abstract}
\noindent We calculate the cross section for $q\bar{q}g$-production in
diffractive DIS with finite fermion masses and zero momentum transfer.
The calculation is done in the leading log(1/$x_{\mathbb P}$)
\mbox{approximation} and is valid for the region of high diffractive masses 
(small $\beta$). We apply our cross section formula to diffractive charm 
production at HERA: in a Monte-Carlo-Simulation of diffractive
$D^{*\pm}$ meson production we include both massive 
$q{\bar q}$- and massive $q\bar{q}g$-production, and we compare with 
preliminary H1 results. After adjusting an infrared cutoff parameter of our 
model which affects the overall normalization of the cross section we find 
reasonable agreement with the data. In particular, the slope of the 
$\beta$-distribution can be reproduced. 
\end{abstract}
\end{center}

\noindent
{\bf 1.} In the process of diffractive deep inelastic scattering,
$\gamma^*+p\rightarrow p+X$, one can separate perturbative and 
non-perturbative contributions by filtering out particular diffractive final 
states. Examples of diffractive states which are perturbatively calculable 
are longitudinal vector particles or final states which
consist of hard jets (and no soft remnant). In the latter 
case the hard scale which allows the use of pQCD is provided by the
large transverse momenta of the jets, and the Pomeron exchange is modelled
by the unintegrated gluon density. Another particularly interesting 
example is diffractive charm production, since the charm quark mass justifies 
pQCD, even for not so large transverse momenta of the outgoing quarks and 
gluons. Calculations for the diffractive production of massless 
open $q\bar q$ states and of massless $q\bar qg$ states have been 
reported in \cite{qq,nikol,gots} and in \cite{qqg}, resp., and a comparison
of diffractive two-jet and three-jet events observed at HERA 
with these calculations has
been presented in \cite{Schilling}. Final states with finite
quark masses have been calculated, so far, only for $q\bar{q}$ production 
\cite{qqm} which is expected to be the dominant final state in the region of
small diffractive masses (large $\beta$). However, as there are recent results from 
measurements of H1 and ZEUS at HERA on diffractive production of (charmed) 
$D^{*\pm}$ mesons, which extend into the small-$\beta$-region, gluon radiation
can certainly not be neglected, and a comparison of the perturbative 
two gluon model with HERA data has not been possible yet. In this letter we report
on a calculation of massive $q\bar{q}g$-production in DIS diffraction, and
we present a comparison of our cross section formula with preliminary H1 data. \\[0.4cm]  
{\bf 2.} We will follow the study of massless $q\bar{q}g$-production 
presented in \cite{qqg}. In particular, we again work in the leading-log $M^2$ 
approximation, which limits the applicability of our results to the small 
$\beta$-region. Fig.\ref{f,notat} shows the notations of the process. As
in \cite{qqg} we restrict ourselves to zero momentum transfer, $t=r^2=0$. 
As usually, $Q^ 2$ denotes the virtuality of the photon, $\sqrt{W^2}$ the 
energy of the photon proton system, $M$ the mass of the diffractive system, 
and $x=Q^2/(Q^2+M^2)$, $y=2pq/2pl$ are the Bjorken scaling variables $x$ and 
$y$ ($l$ is the momentum of the incoming electron). The variable $\beta$ is 
defined as $\beta=Q^2/(Q^2+M^2)$, and it is convenient to introduce 
the momentum fraction of the Pomeron $x_\mathbb P=(Q^2+M^2)/(Q^2+W^2)$. 
The limit we are interested in is defined as: 
\beqn
 Q^2 \ll M^2 \ll W^2. 
\eeqn
\begin{figure}[t] 
 \begin{center}
   \input{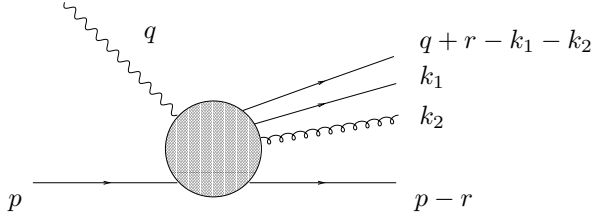} 
   \caption{\label{f,notat}Kinematics of diffractive $q{\bar q}g$ production}
 \end{center}
\end{figure}
We use Sudakov variables $k_i=\alpha_i q' + \beta_i p + k_{i\;t}$ (with
$q'=q+xp$, $k_{i\;t}^2=- {\bf k}_i^2)$, and we express the phase space in 
terms of 
$y$, $Q^2$, $M^2$, $m^2$, $t$, ${\bf k}_1^2$, ${\bf k}_2^2$ with 
$m^2=m_{qq}^2 + {\bf k}_2^2$ ($m_{qq}$ denotes the invariant mass of the 
$q\bar{q}$-subsystem). We obtain the 
following result:
$$\frac{d\sigma_D^{e^-p}}{dy dQ^2 dM^2 dm^2 d^2{\bf k}_1 d^2{\bf k}_2 
dt}_{|t=0}=\frac{\alpha_{em}}{yQ^2\pi} \cdot$$ 
\begin{eqnarray}
\lefteqn{ \cdot\left[ \frac{1+(1-y)^2}{2} \frac{d\sigma^{\gamma^*p}_{D,T+}}{dM^2 dm^2 d^2{\bf k}_1 d^2{\bf k}_2 dt_{|t=0}}-2(1-y) \frac{d\sigma^{\gamma^*p}_{D,T-}}{dM^2 dm^2 d^2{\bf k}_1 d^2{\bf k}_2 dt_{|t=0}}  \right. } \nonumber \\
&&\left.{} +(1-y) \frac{d\sigma^{\gamma^*p}_{D,L}}{dM^2 dm^2 d^2{\bf k}_1 d^2{\bf k}_2 dt_{|t=0}} 
+(2-y)\sqrt{1-y} \frac{d\sigma^{\gamma^*p}_{D,I}}{dM^2 dm^2 d^2{\bf k}_1 d^2{\bf k}_2 dt_{|t=0}} \right] \,,
\label{1}
\end{eqnarray}

\begin{eqnarray}
 \frac{d\sigma^{\gamma^*p}_{D,T+}}{dM^2 dm^2 d^2{\bf k}_1 d^2{\bf k}_2 dt_{|t=0}}&=& \frac{9}{128\pi}\frac{1}{\sqrt{S}(M^2-m^2)m^2} e^2_c \alpha_{em} \alpha^3_s \alpha_1(1-\alpha_1)\cdot \nonumber \\
&& \hspace{1cm} \cdot \left[ \left(\alpha_1^2+(1-\alpha_1)^2 \right)
M_{il} M'_{il}+m_q^2M_lM'_l\right]
\label{2}\\
 \frac{d\sigma^{\gamma^*p}_{D,T-}}{dM^2 dm^2 d^2{\bf k}_1 d^2{\bf k}_2 dt_{|t=0}}&=& \frac{9}{128\pi}\frac{1}{\sqrt{S}(M^2-m^2)m^2} e^2_c \alpha_{em} \alpha^3_s \alpha_1^2(1-\alpha_1)^2 \cdot \nonumber \\
&& \hspace{1cm} \cdot \left[ M_{1l} M'_{1l} - M_{2l}M'_{2l} \right]
\label{3}\\
 \frac{d\sigma^{\gamma^*p}_{D,L}}{dM^2 dm^2 d^2{\bf k}_1 d^2{\bf k}_2 dt_{|t=0}}&=& \frac{9}{128\pi}\frac{1}{\sqrt{S}(M^2-m^2)m^2} e^2_c \alpha_{em} \alpha^3_s 4\alpha_1^3(1-\alpha_1)^3 Q^2 M_l M'_l
\label{4}\\
 \frac{d\sigma^{\gamma^*p}_{D,I}}{dM^2 dm^2 d^2{\bf k}_1 d^2{\bf k}_2 dt_{|t=0}}&=& \frac{9}{128\pi}\frac{1}{\sqrt{S}(M^2-m^2)m^2} e^2_c \alpha_{em} \alpha^3_s \alpha_1^2(1-\alpha_1)^2 (1-2\alpha_1) \cdot \nonumber \\
&& \hspace{1cm} \cdot  \sqrt{Q^2} \left[M_{1l} M'_l + M_l M'_{1l} \right]
\label{5} 
\end{eqnarray}
with
\begin{equation}
S=\left( 1+ \frac{{\bf k}^2_1}{m^2} - \frac{({\bf k}_1 +{\bf
k}_2)^2}{m^2} \right)^2 -4\frac{({\bf k}_1^2 + m_q^2)}{m^2} 
\end{equation}
and 
\[
T_{il}=\left( \frac{{\bf l}+{\bf k}_1+{\bf k}_2}{D({\bf l}+{\bf
k}_1+{\bf k}_2)} +
\frac{{\bf k}_1+{\bf k}_2}{D({\bf k}_1+{\bf k}_2 )} - \frac{{\bf
k}_1-{\bf l}}{D({\bf k}_1-{\bf l})} - \frac{{\bf k}_1}{D({\bf k}_1)} \right)_i
\left(\frac{{\bf l}+{\bf k}_2}{({\bf l}+{\bf k}_2)^2} - \frac{{\bf k}_2}{{\bf k}_2^2} \right)_l 
\]
\begin{equation}
\hspace{3cm}{}+({\bf l} \to -{\bf l})
\end{equation}

\[
T_{l}=\left( \frac{1}{D({\bf l}+{\bf k}_1+{\bf k}_2)} +
\frac{1}{D({\bf k}_1+{\bf k}_2)} - \frac{1}{D({\bf k}_1-{\bf l})} -
\frac{1}{D({\bf k}_1)} \right)
\left(\frac{{\bf l}+{\bf k}_2}{({\bf l}+{\bf k}_2)^2} - \frac{{\bf k}_2}{{\bf k}_2^2} \right)_l 
\]
\begin{equation}
\hspace{3cm}{}+({\bf l} \to -{\bf l}).
\end{equation}
Here
\begin{equation}
M_{il}=\int \frac{d^2{\bf l}}{\pi{\bf l}^2} {\cal F}(x_{\mathbb P},
{\bf l}^2)T_{il}
\label{9}
\end{equation}
and 
\begin{equation}
D({\bf k})=\alpha_1(1-\alpha_1)Q^2 + {\bf k}^2 + m_q^2.
\label{10}
\end{equation}
The function ${\cal F}$ denotes the unintegrated (forward) gluon density 
which is connected with the usual gluon density $g(x,Q^2)$ through:  
\[\int^{Q^2}d{\bf l}^2{\cal F}(x,{\bf l}^2)=xg(x,Q^2).\] 
The parameter $\alpha_1$ is determined by the on-shell conditions for the 
final state particles:
\begin{equation} \alpha_1=\frac{1}{2}\left[ 1+\frac{{\bf k}_1^2}{m^2}-
\frac{({\bf k}_1+{\bf k}_2)^2}{m^2}\pm \sqrt S \right], 
\label{11}
\end{equation}
and it varies between 0 and 1. The values of the
momenta ${\bf k}_1$, ${\bf k}_2$ and of $m^2$ decide which sign in
eq.(\ref{11}) holds. \\[0.4cm]
The quark mass $m_q$ enters the calculations in two places.
First, the phase space of the diffractive system (and so the parameter
$\alpha_1$ and the function $S$) depend upon the quark mass via the
on-shell conditions for the outgoing particles. Secondly, the
propagators of the internal fermion lines are modified by a nonzero
quark mass which leads to changes in the matrixelements. Apart from the
function D({\bf k}), eq.(\ref{10}), which enters all the four
$\gamma^*p$-cross sections, an additional term containing the quark mass 
emerges in $d\sigma^{\gamma^*p}_{D,T+}$, eq.(\ref{2}).\\[0.4cm] 
{\bf 3.} Using our analytic formulae, we have performed a numerical Monte 
Carlo calculation and compared with preliminary H1 data on $D^{*\pm}$ production in
diffractive DIS ($m_{D^*} \approx 2.01 \mbox{GeV}$). 
There are (unpublished) HERA measurements on this process both 
from H1 \cite{h1dstar} and from 
ZEUS \cite{zeusdstar}. Compared to other charmed 
mesons the
$D^{*\pm}$ mesons are easy to reconstruct. This makes them attractive
objects for testing diffractive charm production. Both experiments
made use of the decay channel
\[ D^{*+}\rightarrow D^0\pi^+_{slow} \rightarrow
(K^-\pi^+)\pi^+_{slow} \qquad (\mbox{and c.c.}), \]
which has a branching ratio of 2.63\% \cite{DataG}. The kinematical
regions where ZEUS and H1 have taken their data are slightly different. We
will focus on the comparison with the preliminary H1 data \cite{h1dstar} which have
been collected throughout the years 1995-1997. The amount of
data is still quite poor due to the very rare occurance of the $D^{*}$
in the considered process, and higher statistics will come from new data.
\begin{figure}[h]
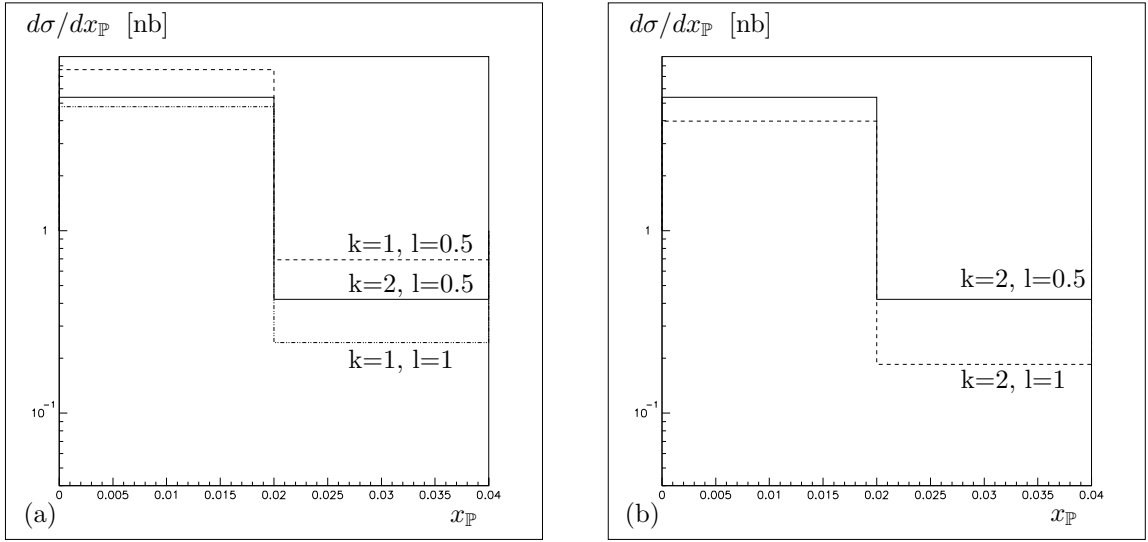

\begin{center}
 \input{xp_param_3.pstex_t} \hspace{0.5cm}
 \input{xp_param_2.pstex_t}
\caption{\label{f,xpxp}d$\sigma / x_{\mathbb P}$  (sum of $c{\bar c}$ and
$c{\bar c}g$ production) at different values of the cut on the gluon
transverse momentum, k=${\bf k}_{2cut}^2$ and of the lower {\bf l} -
integration limit, l=${\bf l}_{min}^2$ (both in GeV$^2$).} 
\end{center}
\end{figure} \\[0.4cm]
We have implemented the ep cross section for diffractive massive $c\bar c$
production both from \cite{qqm} and from our expression eq.(\ref{1}-\ref{5}) 
for the massive $q\bar q g$ production into the Monte-Carlo-Program RAPGAP
2.08 \cite{rapgap}. This program includes hadronization
based on the Lund model. We have added the simulated cross sections of 
$c\bar c$ and the $c\bar cg$ production; the cross sections for $D^{*+}$ 
and $D^{*-}$ production have also been added. $\alpha_s$ was fixed at a value 
of 0.25. We have used the GRV NLO gluon density \cite{GRV} to obtain the 
unintegrated gluon density. The boundary conditions imposed on the
kinematic variables are the same as at H1 \cite{h1dstar}.  The electron
and proton momenta (in the HERA system) are 27.6 GeV and 820 GeV, 
respectively, and we demand:
\[ 0.05 \le y   \le 0.7 \] 
\[ 2    \le Q^2 \le 100 \; \mbox{GeV}^2\]
\[   x_{\mathbb P}<0.04. \]
The momentum transfer of the proton has to be limited, because in the
experimental analysis also events with diffractive excitations of the 
proton have been included in the diffractive cross section: 
\[|t| \le 1 \; \mbox{GeV}^2.\]
Finally, there are two restriction (of more technical origin) on 
the transverse momenta of the $D^*$ mesons and on their pseudorapidity
values $\eta=-\mbox{ln tan}(\theta /2)$ ($\theta$ denotes the angle between 
the momentum of the particle and the direction of the incoming
proton): 
\[ \mbox{in the ep-CMS:} \qquad |\eta(D^{*\pm})|<1.5 \quad \mbox{and} 
\quad p_T(D^{*\pm})>2\;\mbox{GeV}.\] 
In addition to these restrictions which come from the experimental side there
is one further cut on the $c\bar cg$ final state due to non-perturbative 
elements in the calculation. Since the transverse
momentum of the final state gluon defines the scale for the strong
coupling constant we have to impose a lower cutoff ${\bf k}^2_{2cut}$ on
the gluon transverse momentum. If it is choosen to be too large, we loose
on the cross section; at too small values the reliability of perturbation
theory becomes weak. In our calculations we have 
choosen ${\bf k}^2_{2cut}=2\;\mbox{GeV}^2$. Finally, the ${\bf l}$ 
integration of the gluon loop in eq.(\ref{9}) is not defined in the infrared 
region. We therefore choose a nonzero lower integration limit
${\bf l}^2_{min}$ at ${\bf l}^2_{min}=0.5\;\mbox{GeV}^2$. 
In the numerical treatment of the $c\bar c$ production cross section 
neither ${\bf k}^2_{2cut}$ nor ${\bf l}^2_{min}$ are needed.
\begin{figure}[h] 
\begin{center}
 \input{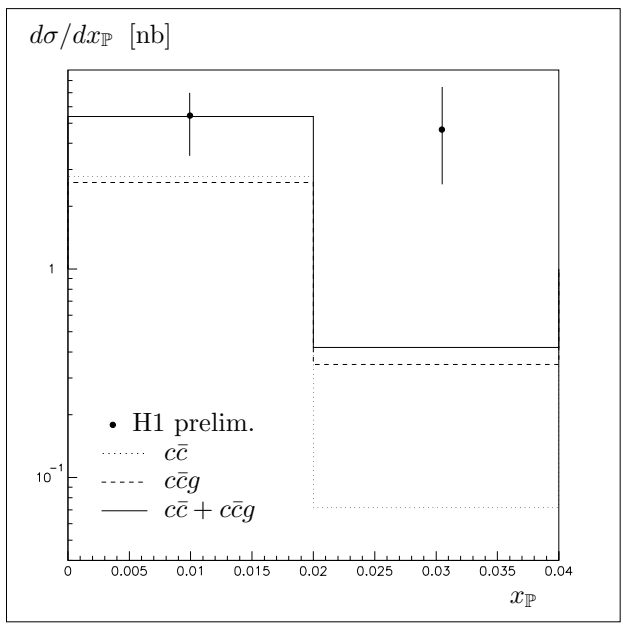}
\caption{\label{f,xp} The cross section of diffractive $D^*$ production
in the 2-gluon-model as a function of $x_\mathbb P$. The data points 
are preliminary H1 data \cite{h1dstar}.}
\end{center}
\end{figure}
\\[0.4cm]
In order to test the sensitivity of our results to the choice of the two 
cutoff parameters 
we have computed $d\sigma /dx_\mathbb P$ for different values. 
Results are shown in fig.\ref{f,xpxp}: the main change is in the overall 
normalization. As expected, lowering any of the two cutoffs leads to an 
increase of the cross section. In our subsequent analysis we have choosen to 
adjust our $x_{\mathbb P}$ distribution ($c \bar c$ and
$c\bar cg$) of the $D^*$ mesons to the data in the lower $x_{\mathbb P}$-bin.
Since there is a certain arbitrariness in this choice, in our comparison
with preliminary H1 data we will concentrate on the shape of the $D^*$ distributions
rather than its normalization.\\[0.4cm]
Turning now to the results of our numerical analysis 
we start with the $x_\mathbb P$ distribution. The contributions due to  
$c\bar cg$ and $c\bar c$ and the sum of both are shown in fig.\ref{f,xp}. 
(as we have said before, the value of the sum of both cross sections 
($c\bar c$ and $c\bar cg$) has been adjusted 
to the data in the lower $x_\mathbb P$-bin). 
In the upper bin the computed cross section is
by a factor of about 10 smaller than the data point. In this
$x_\mathbb P$-region it is expected that, apart from Pomeron 
exchange (which, in our model, is the 2-gluon exchange) 
secondary exchanges have to be included: in a perturbative description 
such an exchange corresponds to $q\bar{q}$-exchange. Since such a contribution
has not yet been included, it is not surprising that the two-gluon model 
undershoots the data.
\begin{figure}[ht] 
\begin{center}
 \input{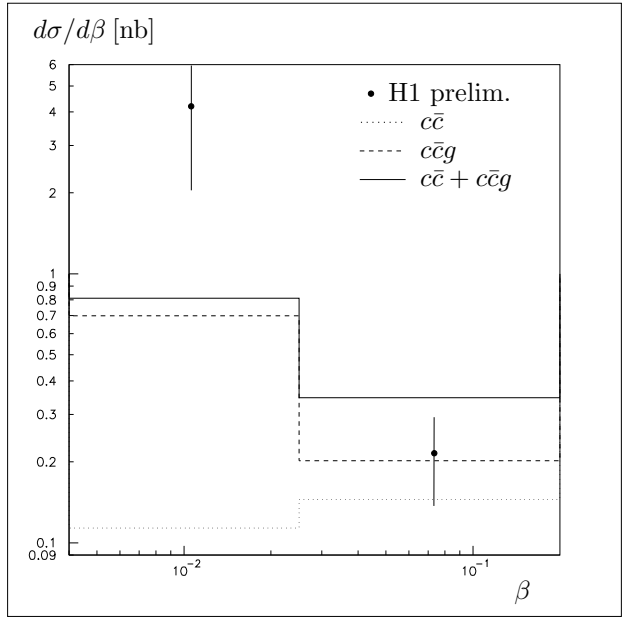}
\caption{\label{f,beta}The cross section of diffractive $D^*$ production
in the 2-gluon-model as a function of $\beta$. Data points are 
preliminary H1 data \cite{h1dstar}.} 
\end{center}
\end{figure}
\\[0.4cm]
The $D^*$ cross section as a function of $\beta$ is shown in
fig.\ref{f,beta}. The measured cross section is rising with decreasing 
$\beta$. This property is not reproduced by the $c\bar c$ channel
alone. Only when both $c\bar c$ and $c\bar cg$ production
are included the $\beta$ distribution of the two gluon model increases 
with decreasing $\beta$. This fact is nearly independent of the choice of the 
cutoffs ${\bf l}^2_{min}$ and ${\bf k}^2_{cut}$ which mainly affect the 
overall strength of the cross section. Therefore the $c\bar cg$
contribution really improves the description of the $\beta$
distribution. The calculated sum of the cross sections overshoots the data 
in the upper $\beta$-bin and still remains below data in the lower
$\beta$-bin.
A reason for the disagreement in the lower $\beta$-bin could be the radiation 
of more than one gluon which might be present in the data but is not included in 
our model. Since in the small $\beta$-region $x_\mathbb P$ gets larger, the
disagreement could also be due to our neglect of secondary exchanges.\\[0.4cm]
Two further distributions are shown in fig.\ref{f,q2_y}. In contrast
to the $c\bar c$ contributions alone the full cross sections
in both bins lie close to the data points
(except for the low-$y$ point in fig.\ref{f,q2_y}b even within the error bars of the
data points). The ratio of the data values in the lower and upper $Q^2$ bins
is about 15. For $c\bar c$ alone the model gives the value 8.3; 
when the $c\bar cg$ contribution is included it goes 
up to 10.7. Also in the $y$ distribution (fig.\ref{f,q2_y}b) the agreement
of the sum of both cross sections with data is better than for 
$c\bar c$ alone.
\begin{figure}[h]
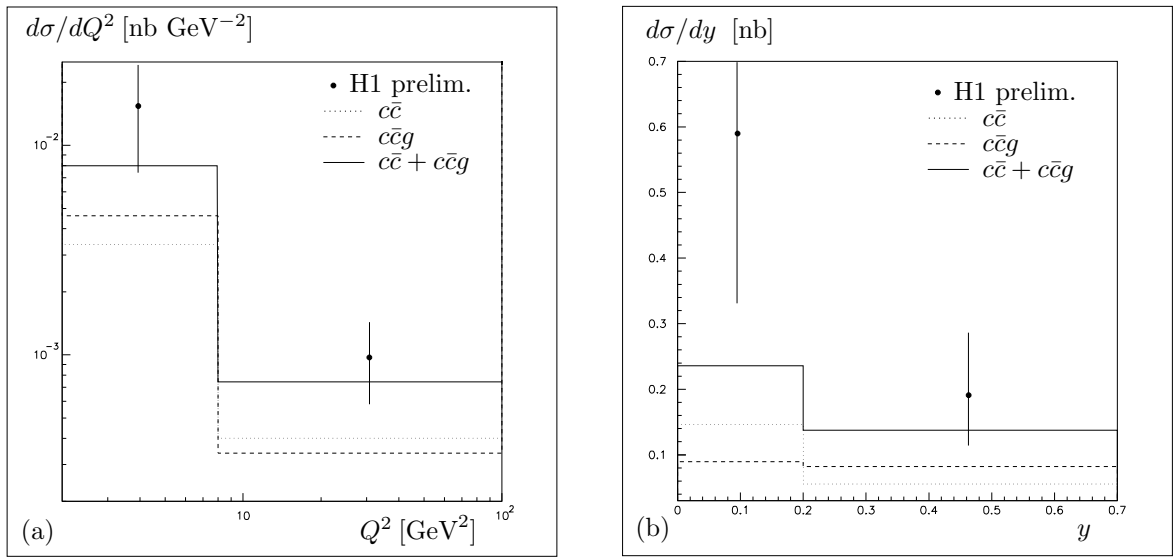
 
\begin{center}
 \input{q2.pstex_t} \hspace{0.5cm}
 \input{y.pstex_t}
\caption{\label{f,q2_y}The cross section of diffractive $D^*$ production
in the 2-gluon-model as a function of $Q^2$ and as a function of $y$. 
Data points are the H1 preliminary data \cite{h1dstar}.} 
\end{center}
\end{figure}

\noindent
{\bf 4.}
In this letter we have analysed, within the perturbative two-gluon model, 
DIS diffractive charm production (production of $D^{*\pm}$ mesons). 
Compared to an earlier attempt where
only $c\bar{c}$ production had been included in the theoretical analysis
the present analysis contains, as the new ingredient, also (massive)
$c\bar{c}g$ production and leads to a considerable improvement in the 
agreement with experimental data.\\[0.4cm]
Despite this encouraging success, several improvements in the theoretical
part should be made. First, the cross section formula for     
$c\bar{c}g$ production has been calculated in the 
leading log-$M^2$ approximation; an improvement which extends the applicability
down to small-$M^2$ values would be very desirable. For consistency reasons,
one then will need a NLO-calculation of $c\bar{c}$ production. 
Next, our comparison with data indicates the need of $c\bar{c}gg$ final states:
such an extension (at least in the leading log-$M^2$ approximation)
should be fairly straightforward. Finallly, the region of   
$x_\mathbb P>0.02$ seems to require secondary exchanges which, in the framework  
of perturbative QCD, should be modelled by $q\bar{q}$ exchange.\\[0.4cm]
A successful test of the two-gluon model in DIS Diffraction, apart form 
providing a description of charm or jet production at HERA, is also of 
general theoretical interest: the cross section formula for diffractive 
$q\bar{q}+ng$ production contains the perturbative triple Pomeron vertex 
which is expected to play a vital role in the unitarization of the BFKL
approximation. 
It has been calculated both analytically and numerically, and these
calculatutions can be tested experimentally in DIS diffraction dissociation.   


\begin{thebibliography}{99}
\bibitem{qq} J.Bartels, H.Lotter, M.W\"usthoff, Phys. Lett.{\bf B 379}
(1996) 239; ERRATUM-ibid. {\bf B382} (1996) 449.
\bibitem{nikol} N.Nikolaev, B.G. Zakharov, Z.Phys.{\bf C53} (1992) 331.
\bibitem{gots} E.Gotsman, E.Levin, U.Maor, Nucl. Phys. {\bf B493}
(1997) 354.
\bibitem{qqg} J.Bartels, H.Jung, M.W\"ustoff, Eur.Phys.J. {\bf
C11}(1999) 111.
\bibitem{Schilling} K.P.Schilling (H1 Collaboration), Diffractive Dijet and 3-jet 
Electroproduction at HERA;\\ 
proceedings from DIS 2000, Liverpool.
\bibitem{qqm} H.Lotter, Phys. Lett. {\bf B406} (1997) 171. 
\bibitem{h1dstar} H1 Collaboration, Measurement of the Production of 
$D^{*\pm}$Mesons in Deep Inelastic Interactions at HERA;
Contributed Paper 157ag, International Europhysics Conference
on High Energy Physics (HEP99), Tampere, Finland, 1999; S. Hengstmann,
Proc. 7th International Workshop on Deep-Inelastic Scattering and QCD
(DIS99) Zeuthen, 1999., July 1999.
\bibitem{zeusdstar}ZEUS Collaboration, Study of $D^{*\pm}$ Meson Production in Diffractive Deep 
Inelastic ep Scattering at HERA;
EPS99 contribution from Diffractive and VM group
[EPS 527, plenary: 3 5 18, parallel 3 5]  
\bibitem{DataG} C.Caso et al. (Particle Data Group),
Eur.Phys.J. {\bf C3} (1998) 1.
\bibitem{rapgap} H.Jung, Comp. Phys. Comm. {\bf 86} (1995) 147. \\
 H.Jung, The RAPGAP Monte Carlo for Deep Inelastic Scattering, version
2.08,\\ Lund University, 2000,\\  
\verb+http://www-h1.desy.de/~jung/rapgap.html+.
\bibitem{GRV} M.Gl\"uck, E.Reya, A.Vogt, Z.Phys. {\bf C 67} (1995) 443.
\end{thebibliography}
\end{document}